\documentstyle[11pt,paspconf,epsf]{article}

\begin{document}

\title{Foregrounds and experiments below 33~GHz}
\author{A. W. Jones}
\affil{Astrophysics Group, Cavendish Laboratory, Madingley Road,
Cambridge, CB3 0HE, UK}

\begin{abstract}
In this paper I report on the recent results obtained from the
Tenerife switched beam (10~GHz to 33~GHz) experiment
and Jodrell Bank 5~GHz interferometer and their
predictions for foreground levels. The full data analysis and results
will be presented elsewhere. It is found that free-free emission
dominates above 10~GHz, whereas synchrotron emission dominates below
5~GHz. 
\end{abstract}


\begin{center}
{\em In the beginning there was only darkness, dust and water. The darkness
was thicker in some places than in others. In one place it was so thick
that it made man. The man walked through the darkness. After a while, he
began to think.}
\end{center}

\begin{center}
Creation myth from the Pima tribe in Arizona
\end{center}

\section{Introduction}

One of the biggest problems in improving the results from current
Cosmic Microwave Background (CMB) experiments is that of
foreground contamination. Without prior knowledge of the spatial and
frequency spectra of foregrounds it is impossible to say for certain
which components present in the data can be attributed to the CMB or to
Galactic or extra-Galactic foregrounds. The only way to overcome this
problem is to use multi-frequency information to constrain the various
foregrounds and derive their properties. This paper will try to put
some extra constraints on the dominant foregrounds at the lower
frequency range (5~GHz to 30~GHz) that is of interest to CMB
astronomy. 

\section{The Tenerife experiments}

[Project collaborators: {\em C.M. Gutierrez, A.W. Jones, R.D. Davies,
R. Watson, S. Melhuish, A.N. Lasenby, R. Rebolo, R. Hoyland,
P. Mukherjee, A. de Oliveira-Costa}]

The Tenerife switched beam radiometers have been operating since 1984
at the Teide Observatory in Tenerife. There are three instruments
operating at 10~GHz, 15~GHz and 33~GHz. The beams have a full width half
maximum (FWHM) of about $5\deg$ and a triple beam pattern is obtained
by switching two beams in right ascension (RA) by $8.1\deg$. More
details about the experiments can be found in Davies {\em et al.}
(1996) and the most
recent results are to be published in Gutierrez {\em et al.} (1999). 

\begin{figure}
\plotfiddle{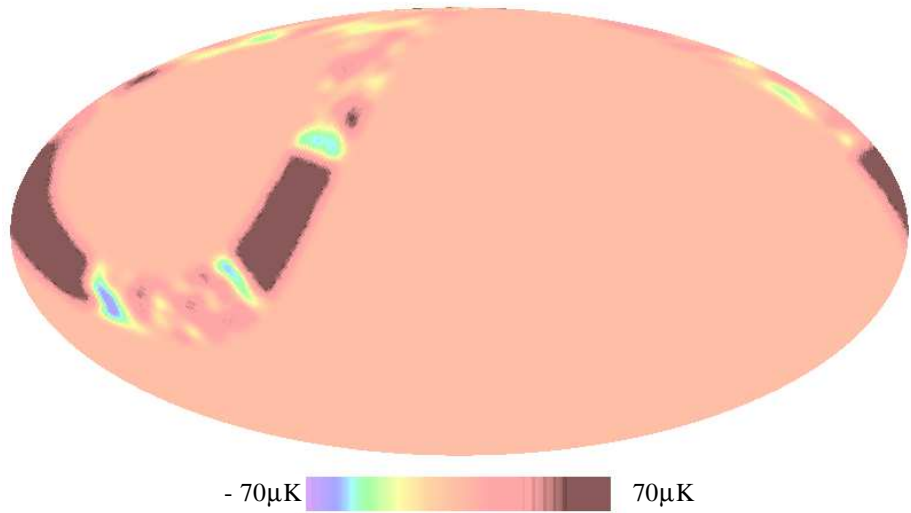}{6cm}{270}{45}{45}{-180}{260}
\caption{The Tenerife 15~GHz data shown in Mollwiede projection (so
that the Galactic plane crossing is along the equator).}
\label{fig1}
\end{figure}

The data from the experiments are analysed in two different
ways. Statistical techniques (for example, the likelihood function) 
are used to obtain constraints on cosmological 
parameters from the raw data (after calibration) and maximum entropy
(Jones {\em et al.} 1998) is used to construct maps of the sky at $5\deg$
resolution. Figure~\ref{fig1} shows the 15~GHz Tenerife result
produced by maximum entropy in the Mollwiede projection. The average
sensitivities of the three frequencies are $48\mu$K, $19\mu$K and
$30\mu$K per beam for the 10, 15 and 33~GHz instruments
respectively. Additional information on the point sources (data taken by the
Michigan monitoring program; Aller \& Aller 1997) has
been taken simultaneously over a 10 year period so accurate subtraction
of this foreground is possible and has been carried out here prior to
the Galactic foreground analysis. 

\subsection{Cross-correlation with existing templates}

The first attempt at calculating the level of Galactic foreground in
the Tenerife data is to calculate the cross-correlation with other
data which is known to be dominated by the foregrounds.
We have cross-correlated the raw data from the
10~GHz and 15~GHz (the 33~GHz does not have enough sky coverage) with
templates from the DIRBE experiment (1-3~THz) to test for dust or
correlated free-free emission and with the two low frequency surveys
at 408~MHz (Haslam, Salter, Stoffel \& Wilson 1982) and 1420~MHz (Reich \& Reich
1988). Table~\ref{table1} summarises the results from this analysis. 

As can be seen from the table no significant correlation was found between
the low, or high, frequency data and the 15~GHz Tenerife data. It is
therefore possible to say that the templates for free-free and
synchrotron do not contribute greatly to the 15~GHz data and that most
of this signal is likely to be CMB in origin. However, at 10~GHz a
$\sim 3$ sigma detection was found with both the 408~MHz and 1420~MHz
surveys with a signal corresponding to about half 
of the total signal
($\Delta T_{10}=29^{+10}_{-10}\mu$K; Gutierrez {\em et al.} 1999)
present in the 10~GHz data. Therefore, it is very likely that
free-free or synchrotron emission is a significant foreground in this
channel. However, care must be taken as artefacts in both of the low
frequency maps may also be contributing to the correlation and so
further analysis must be carried out. 

\begin{table}
\begin{center}
\begin{tabular}{|c|c|c|c|c|} \hline
& \multicolumn{2}{|c|}{10~GHz} & \multicolumn{2}{|c|}{15~GHz} \\ \cline{2-5}
Template & Significance & Signal ($\mu$K) & Significance & Signal
($\mu$K) \\ \hline
DIRBE08 & 0.93 & 8.5  & 1.44 & 7.8 \\
DIRBE09 & 0.19 & 1.7  & 0.35 & 1.7 \\
DIRBE10 & 1.06 & 9.5  & 0.96 & 5.1 \\
HASLAM  & 3.74 & 32.7 & -0.07 & -0.4 \\
R\&R    & 2.32 & 22.4 & 0.53 & 3.0 \\ \hline
\end{tabular}
\end{center}
\caption{Results from the cross-correlation of the 10~GHz and 15~GHz
Tenerife data with the Dirbe 100 (DIRBE08), 140 (DIRBE09) and
240$\mu$m (DIRBE10), Haslam 408~MHz and Reich \& Reich 1420~MHz
surveys. The significance in standard deviations is shown and the
corresponding signal contributing to the data is shown in $\mu$K.}
\label{table1}
\end{table}

\subsection{Joint likelihood analysis}

Previous likelihood analyses have been performed on the Tenerife data
but always using the frequencies as separate channels (either assuming
that all the signal present is either CMB or Galactic). However, if we
perform a joint likelihood analysis on the overlap region between the 10~GHz
and 15~GHz data assuming some spectral index for the Galactic signal
then it will be possible to extract CMB and Galactic results
simultaneously. We assume that each pixel has a temperature which can
be represented by 

\begin{equation}
\Delta T^{10,15}=\Delta T_{CMB} + \Delta T_{Gal}^{10,15}
\label{eq1}
\end{equation}

\noindent
where 

\begin{equation}
\Delta T_{Gal}^{15} = \Delta T_{Gal}^{10} \times(15/10)^{-\beta}
\label{eq2}
\end{equation}

\begin{figure}
\plotone{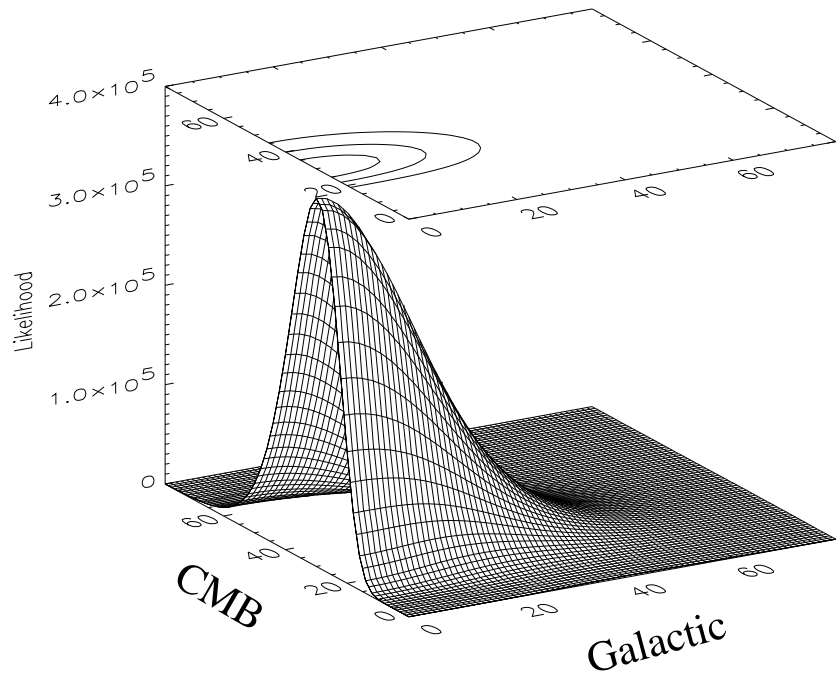}
\caption{The 2-D likelihood function for the Galactic and CMB
contribution to the 10 and 15~GHz Tenerife experiments. The
temperatures are shown in $\mu$K.}
\label{fig2}
\end{figure}

\noindent
Figure~\ref{fig2} shows the two dimensional likelihood function for
the joint analysis of the 10 and 15~GHz data as a function of the
temperature of the CMB and Galactic signal at 10~GHz. As can be seen
there is no firm detection of the Galactic signal, only an upper limit
($\Delta T_{Gal}^{10}<28\mu$K at 68\% confidence limits)
but the CMB is detected with a signal which corresponds to 
$Q_{rms-ps}=20^{+10}_{-7}\mu$K (assuming a Harrison-Zeldovich
spectrum). This is in agreement with the COBE result of 
$Q_{rms-ps}=18\pm 1.6\mu$K (Bennett {\em et al}. 1996). 

\subsection{Foreground map making}

Another possible analysis technique is to use the frequency
information on a pixel by pixel basis to perform a multi-frequency
decomposition of the data. To put more constraints on the data we have
incorporated the common region of sky that was observed using the COBE
satellite. We allowed the spectral index of the Galactic channel to
vary and found that the most likely value was consistent with
free-free emission. Multi-frequency maximum entropy (Hobson, Jones,
Lasenby \& Bouchet 1998) was used
(at these angular scales both the signal and noise are 
Gaussian so the maximum entropy
method reduces to the simple Wiener filter) to analyse the data. 
See Jones {\em et al.} (1999a) 
for a detailed study of the results. 

\section{The Jodrell Bank interferometer}

[Project collaborators: {\em R.D. Davies, S. Melhuish, A.W. Jones,
G. Giardino, A. Wilkinson, H. Asareh, A.N. Lasenby}]

A 5~GHz interferometer has been operating at Jodrell Bank since
1991. It has a primary beam of $8\deg$ FWHM and data has been taken on
two baselines corresponding to angular sensitivities of 
$\ell=71\pm 13$ (the narrow spacing) and $\ell=184\pm 20$ (the wide
spacing). It is not a tracking interferometer so the angular
resolution in Declination is different from that in RA and this has to
be taken into account in the analysis. For more details on the
instrument see Melhuish {\em et al.} (1997) 
and for more details on the data and subsequent
analysis see Giardino {\em et al.} (1999), Asareh {\em et al.} (1999) 
and Jones {\em et al.} (1999b). Again, maximum entropy was used to
produce the maps shown here. The sensitivity per beam for the two
baselines is $20\mu$K. 

\begin{figure}
\plotfiddle{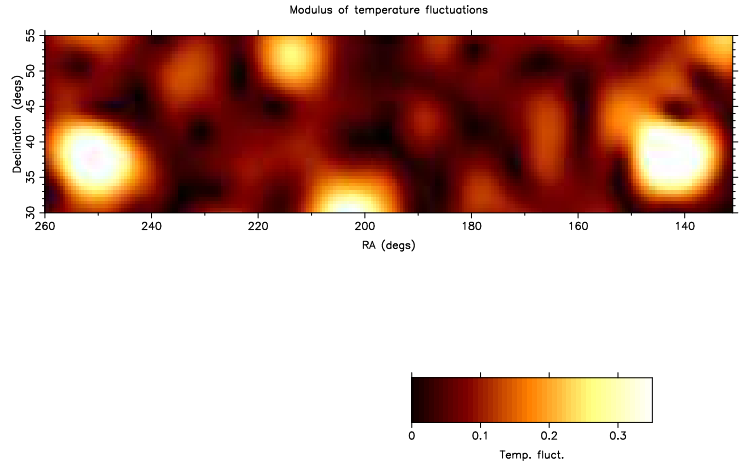}{6cm}{270}{60}{60}{-200}{310}
\caption{The data from the 5~GHz wide spacing interferometer showing
the point source contribution at that frequency. The temperature is
shown in mK.} 
\label{fig3}
\end{figure}

\begin{figure}
\plotfiddle{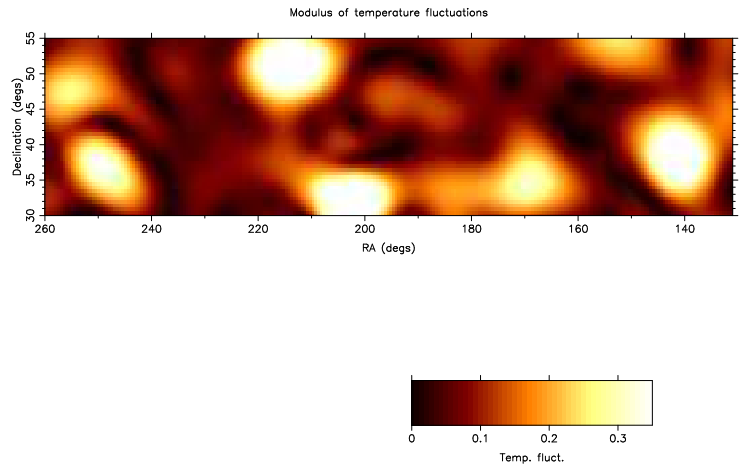}{6cm}{270}{60}{60}{-200}{310}
\caption{The data from the 5~GHz narrow spacing interferometer which
shows a clear excess of sources above the point sources
(Figure~\ref{fig3}). The temperature is shown in mK.}
\label{fig4}
\end{figure}

The data from the wide spacing configuration is shown in
Figure~\ref{fig3}. As can be seen the four main point sources in this
region are easily identifiable. Comparing this map to that produced by
the narrow spacing data (Figure~\ref{fig4}) an excess signal is
clearly seen. This is due to Galactic features and by performing a
multi-frequency analysis with the 408~MHz and 1420~MHz surveys (see
Jones {\em et al.} 1999b) 
it was found that the Galactic spectral index is $\beta=2.9\pm
0.3$, consistent with a synchrotron origin. 

\section{Discussion}

\begin{figure}
\plotfiddle{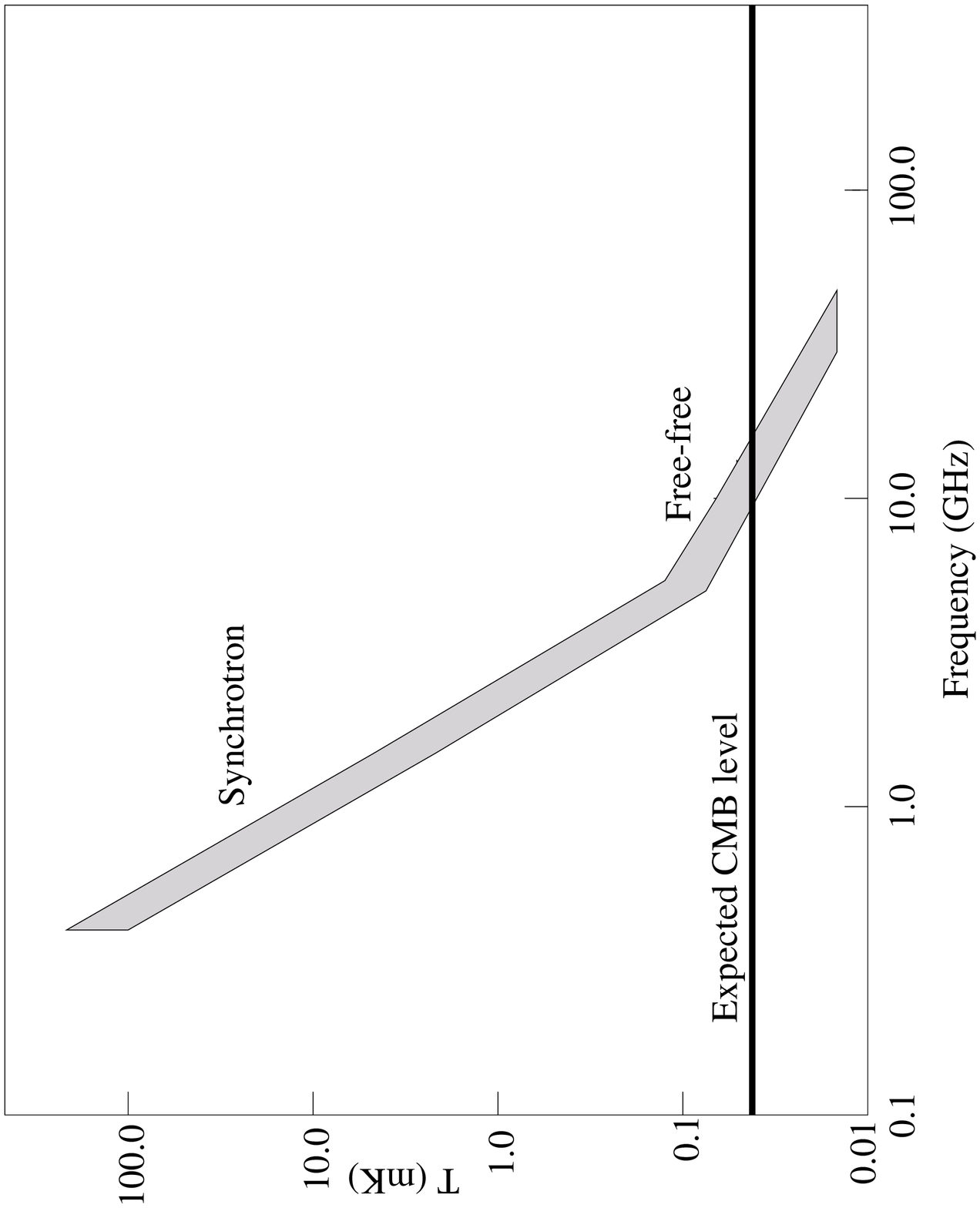}{5cm}{270}{40}{40}{-190}{220}
\caption{Estimate of the dominant foreground level from the results
presented in this paper. The levels shown are for a $5\deg$ FWHM experiment.}
\label{fig5}
\end{figure}

It is found that there is significant Galactic contamination even in
the relatively clean area away from the Galactic plane that has been
observed by the Jodrell Bank interferometer and the Tenerife
beam-switching experiments. Analysis using the new
multi-frequency maximum entropy methods showed that synchrotron
emission dominates below 5~GHz while free-free emission dominates
above 10~GHz. Figure~\ref{fig5} summarises the expected levels of the
low frequency foregrounds for a $5\deg$ FWHM experiment. It is therefore 
very difficult to use the low frequency templates to predict the
position of Galactic foregrounds at higher frequencies. It is
important to note that the predictions 
presented here are most likely lower limits
for the Galactic emission in other regions of the sky.

\acknowledgments

We are grateful to all those involved in the Tenerife and Jodrell Bank
experiments and their continued perseverance in correct art of data analysis.
The author acknowledges King's College, Cambridge, for support in the
form of a Research Fellowship.

\end{document}